\documentclass[preprint2]{aastex}
\usepackage{amsmath}
\usepackage{color}
\usepackage{amssymb}
\usepackage[normalem]{ulem}
\usepackage{array}
\usepackage{soul}
\newcolumntype{L}{>{\centering\arraybackslash}m{5cm}}
\usepackage{natbib}
\usepackage{footnote}

\usepackage{booktabs}
\usepackage{tabularx}

\begin{document}


\title{Characterization of \emph{Kepler}-91b and the Investigation of a Potential Trojan Companion Using EXONEST}


\author{Ben Placek and Kevin H. Knuth \altaffilmark{1}}
\affil{Physics Department, University at Albany (SUNY),
    Albany, NY 12222}

\author{Daniel Angerhausen \altaffilmark{2}}
\affil{Department of Physics, Applied Physics, and Astronomy, Rensselear Polytechnic Institute, Troy, NY 12180 \\ NASA Goddard Space Flight Center, Exoplanets \& Stellar Astrophysics Laboratory, Code 667, Greenbelt, MD 20771}

\author{Jon M. Jenkins}
\affil{NASA Ames Research Center, Moffett Field, CA 94035}


\altaffiltext{1}{Department of Informatics, University at Albany (SUNY), Albany, NY 12222}
\altaffiltext{2}{NASA Postdoctoral Program Fellow, Goddard Space Flight Center, Greenbelt, MD, 20771}



\begin{abstract}
Presented here is an independent re-analysis of the \emph{Kepler} light curve of Kepler-91 (KIC 8219268).  Using the EXONEST software package, which provides both Bayesian parameter estimation and Bayesian model testing, we were able to re-confirm the planetary nature of \emph{Kepler}-91b. In addition to the primary and secondary eclipses of \emph{Kepler}-91b, a third dimming event appears to occur approximately $60^o$ away (in phase) from the secondary eclipse, leading to the hypothesis that a Trojan planet may be located at the L4 or L5 Lagrange points.  Here, we present a comprehensive investigation of four possibilities to explain the observed dimming event using all available photometric data from the \emph{Kepler} Space Telescope, recently obtained radial velocity measurements, and N-body simulations.  We find that the photometric model describing \emph{Kepler}-91b and a Trojan planet is highly favored over the model involving \emph{Kepler}-91b alone.  However, it predicts an unphysically high temperature for the Trojan companion, leading to the conclusion that the extra dimming event is likely a false-postive.  
\end{abstract}

\keywords{Bayesian Methods, Exoplanet Detection, Model Selection, Bayesian Evidence, Trojan, \emph{Kepler}-91, KIC 8219268}

\section{Introduction}

\begin{table*}
\centering
\small
\begin{tabular}{l c}
\hline
Stellar Parameters  From Literature& \\
\hline
Mass, $(M_\odot)$ & $1.31 \pm 0.10$ \\
Radius, $(R_\odot)$ & $ 6.30 \pm 0.16$\\
Effective Temperature, (K) & $4550 \pm 75$ \\
Surface Gravity, $\log g_\star, (c.g.s.)$ & $2.953 \pm 0.007$\\
Gravity Darkening Exponent & $0.733$ \\
Lin. Limb Darkening Coeff. & $0.549$ \\
Quad. Limb Darkening Coeff. ($u_1,u_2$) &$(0.69, 0.05)$ \\
\emph{Kepler} Magnitude & 12.495 \\
\hline
Estimated Planetary Parameters & \\
\hline
Mass, $(M_J)$ & $1.09 \pm 0.20$ \\
Radius, $(R_J)$ & $1.384^{+0.011}_{-0.054}$ \\
Density, $(\rho_J)$ & $0.33^{+0.33}_{-0.05}$ \\
Period, $(d)$	& $6.24650 \pm 0.000082$ \\
Inclination, $(deg)$ & $68.5^{+1.0}_{-1.6}$ \\
Eccentricity & $0.066^{+0.013}_{-0.017}$\\
$a/R_\star$ & $2.32^{+0.07}_{-0.22}$ \\
Equilibrium Temperature, (K) & $2460^{+120}_{40}$ \\ 
\hline
\end{tabular}
\caption{Literature stellar parameters for \emph{Kepler}-91 (KIC 8219268) and estimated planetary parameters for \emph{Kepler}-91b from \citet{Lillo-Box+etal:2013, Lillo-Box+etal:2014}. }
\label{tbl:acceptedvalues}
\end{table*}

\emph{Kepler}-91 (KIC 8219268) is a red giant branch star in a late stage of stellar evolution. It is orbited by at least one companion (Kepler-91b), which was determined to be of planetary nature by \citet{Lillo-Box+etal:2013}.  The planetary status has been refuted by two independent groups \citep*{Esteves+etal:2013, Sliski&Kipping:2014}, but has since been confirmed by radial velocity measurements \citep{Barclay+etal:2014, Lillo-Box+etal:2014}.  \emph{Kepler}-91b previously has been estimated to be a transiting Jupiter-mass planet ($M_p = 0.88^{+0.17}_{-0.33} M_J$) that orbits its red giant host star at one of the shortest known orbital distances of just $a = 2.32^{+0.07}_{-0.22}R_\star$ with an orbital period of $T = 6.24650 \pm 0.000082$ days \citep{Batalha+etal:2013, Lillo-Box+etal:2013, Lillo-Box+etal:2014} (Literature values in Table \ref{tbl:acceptedvalues}).  It orbits its host star in a slightly eccentric orbit of $e = 0.066$ inclined by $68.5^o$ with respect to the plane of the sky.  Being so close to such a giant star ($R_\star = 6.30 \pm 0.16 R_\odot$), even at this low inclination the planet is still observed to transit \emph{Kepler}-91.  In addition to the primary transit and secondary eclipse corresponding to \emph{Kepler}-91b, \citet{Lillo-Box+etal:2013} observed what appears to be a third occultation that occurs approximately $60^o$ in orbital phase away from the secondary eclipse.  They describe four potential hypotheses explaining this dimming event: (1) a transiting Trojan planet located at either the L4 or L5 Lagrange point, (2) a second more distant planet in a resonant, non-coplanar orbit, (3) the presence of a large transiting exomoon in a resonant orbit around \emph{Kepler}-91b, (4) instrumental effects associated with the \emph{Kepler} pipeline or stellar variability.

Previous theoretical studies of exo-Trojans include numerical simulations to investigate the stability of such worlds \citep{Dvorak+etal:2004, Schwarz+etal:2007, Schwarz+etal:2009}, as well as the effect that a Trojan planet would have on the observed transit times of the primary planet (Transit Timing Variations) \citep{Ford&Gaudi:2006, Ford&Holman:2007, Madhusudhan&Winn:2009}, and a search for Trojans in binary systems based on anomalies in the light curves associated with the correct phase expected from a Trojan \citep{Caton+etal:2001}.  Despite these studies, the definitive discovery of such a planet remains elusive.

Presented here is a brief discussion of the four possible hypotheses intended to describe the observed flux, and an analysis of approximately $1500$ days worth of \emph{Kepler} observations and published radial velocity measurements of the \emph{Kepler}-91 system.  Three planetary models are applied to the \emph{Kepler}-91 light curve: two, which model the photometric and radial velocity signals from a single planet, \emph{Kepler}-91b, in the case of both circular and eccentric orbits, and another that models the photometric and radial velocity signals from \emph{Kepler}-91b and an exo-Trojan in an eccentric orbit.  The analysis was performed using the EXONEST software package \citep{Placek+etal:2014}, which relies on Bayesian model selection to statistically test between competing models, thus determining which model best describes a given dataset.  Since at present, there are no confirmed instances of exo-Trojans, this would be the first photometric evidence of an entirely new class of exoplanet.  Furthermore, its existence would provide a unique opportunity to study the effect of stellar flux on two different planets in identical orbits.

\subsection{The Trojan Hypothesis}\label{trojan hypothesis}
\begin{figure}[h!]
\centering
\includegraphics[width = 7cm]{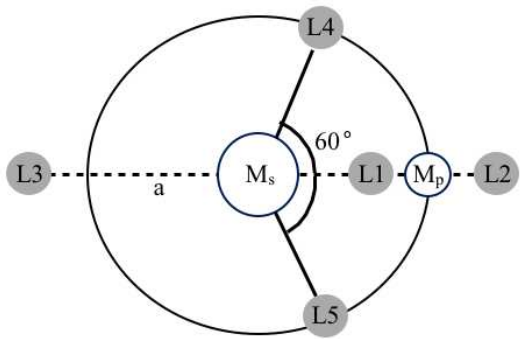}
\caption{Schematic of Lagrange points in the restricted three-body problem.  The L4 and L5 Lagrange points are located at the same orbital distance from the star, but lag or lead the primary planet by $60^o$. }
\label{fig:lagrange}
\end{figure}

In the restricted three-body problem, there are five equilibrium points known as Lagrange points.  The points L1, L2, and L3 lie on the line connecting the primary planet to the host star, whereas L4 and L5 are positioned $60^o$ in front of, or behind, the primary planet in its own orbit as shown in Figure \ref{fig:lagrange}.  Any object located at the L4 or L5 Lagrange point will appear stationary (neglecting librations) in a co-rotating reference frame.  That is, it will orbit the host star with the same period as that of the primary planet.  These orbits at L4 and L5 are stable provided that the host star is greater than 24.96 times the mass of the primary planet, whereas L1, L2, and L3 are unstable saddle points of the gravitational effective potential.

 A massive object located at the L4 or L5 Lagrange point will undergo librational motion.  The object will display long-period librations about the Lagrange point with period
\begin{equation} \label{long libration}
T_{\rm long} = \frac{T_J}{\sqrt{27/4\mu}},
\end{equation} 
where $T_J$ is the orbital period of the primary planet, and $\mu = M_p/(M_\star+M_p)$, as well as epicyclic librations with period
\begin{equation}\label{short libration}
T_{\rm short} = \frac{T_J}{\sqrt{1-\frac{27}{8}\mu}}
\end{equation}

The phase difference of $60^o$ between the secondary eclipse of \emph{Kepler}-91b and the third dimming event observed in the \emph{Kepler}-91 light curve makes the possibility of a Trojan body very enticing.  However, it should be noted that this dimming event would have to be the result of the secondary eclipse of the Trojan --- not the primary transit.  If it were a primary transit, then the second body would be out of phase from \emph{Kepler}-91b by $120^o$, and thus not be a Trojan. 
The fact that this signal appears in the light curve folded at the orbital period of \emph{Kepler}-91b, implies that it either occurs at the same period as \emph{Kepler}-91b, or at an integer value of that period.

\subsection[]{The Distant Planet Hypothesis}\label{dist_plan}

As mentioned in the previous section, this signal could be the transit of a more distant planet in a non-coplanar, resonant orbit to that of \emph{Kepler}-91b.  The low signal to noise of the raw \emph{Kepler} light curve of \emph{Kepler}-91 makes it difficult to simulate such a complex situation, however by considering the light curve folded on the period of \emph{Kepler}-91b, one may be able to determine the likelihood of such a configuration. 
If the occultation is due to the presence of a more distant resonant planet, one would expect differences between the occultations occurring at odd, or even integrals of the orbital period of \emph{Kepler}-91b. 

\begin{figure*}
\centering
\includegraphics[width=16cm]{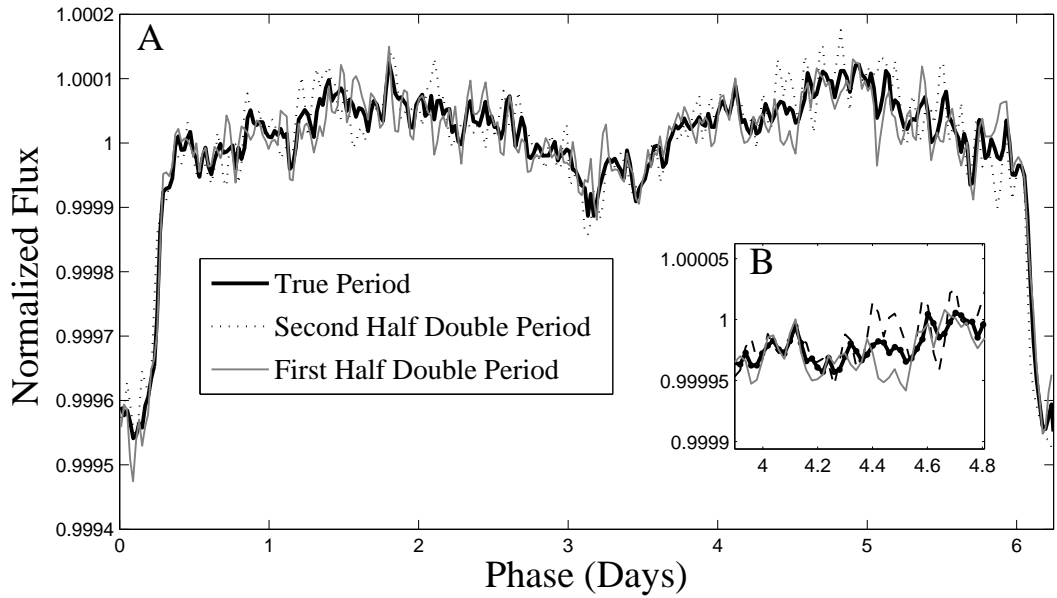}

\caption{Phase folded bin-averaged \emph{Kepler} light curve of \emph{Kepler}-91b.  The solid black curve represents the light curve folded on the true period of $6.24650$ days, while the solid gray and black dashed curves represent the first and second halves of the light curve folded on twice the true period. Inset B shows the discussed extra dimming.}
\label{fig:half periods}
\end{figure*}

Figure \ref{fig:half periods}, shows the binned time series for \emph{Kepler}-91b, folded at the accepted orbital period of \emph{Kepler}-91b (solid black curve), and twice that of \emph{Kepler}-91b (solid gray and dashed black curves). One can clearly see differences between the third dimming event occurring during the first and second halves of the double-period.  While there appears to still be an occultation occurring, there is an odd feature toward one side of the dimming. During the first half of the double-period, there is an increase in the flux while during the second half of the double-period there is a decrease occurring at the same orbital phase, which can be seen more clearly in Figure \ref{fig:half periods}B.  There are however significant odd/even effects for the primary eclipse of \emph{Kepler}-91b as well, implying that this hypothesis may not be the correct explanation.  These odd/even effects will be discussed in more detail in Section \ref{sec:even_odd}

\subsection[]{The Exomoon Hypothesis}\label{exo_moon}

A sizable exomoon in a resonant orbit may also be the cause of this third dimming event. The moon would need to appear to transit \emph{Kepler}-91b $60^o$ in phase before or after its secondary eclipse of the host star.  In addition, this exomoon would need to have an orbital period that is the same as, or an integer multiple of that associated with \emph{Kepler}-91b.  
 This exomoon hypothesis and the stability of such a configuration is discussed in detail in Section \ref{Exomoon Stability Section}.   

\subsection[]{Instrumental Effects and Stellar Variability}\label{instru}
The \emph{Kepler} science pipeline component Presearch Data Conditioning (PDC) identifies and removes systematic errors that are highly correlated across the ensemble of target stars on each CCD readout channel. It also identifies outliers and transients associated with radiation damage to the CCDs at the pixel level, called sudden pixel sensitivity dropouts.  Global instrumental signatures are identified via singular value decomposition (SVD) of the quiet stars on each CCD readout channel. PDC-MAP applies a Bayesian maximum a posteriori (MAP) approach \citep{Stumpe+etal:2014, Smith+etal:2012} to constrain the fit coefficients for the SVD-identified co-trending basis vectors that are then projected out of each stellar light curve. This data reduction step can coincidentally remove stellar variations correlated with the instrumental signatures, especially on timescales longer than 20 days and can in some cases, introduce systematic signatures into individual light curves. However, it is highly unlikely that PDC-MAP introduced consistent brightening and dimming events at the period of \emph{Kepler}-91b, as such signatures are not evident in the co-trending basis vectors which are formulated for each individual observing segment or quarter.  Moreover, the transits and secondary occultation of \emph{Kepler}-91b and the deep secondary occultation of the candidate Trojan planet are readily identifiable in the simple aperture photometry (PA-SAP) produced by the \emph{Kepler} science pipeline, in addition to the PDC-MAP light curves analyzed in this paper \citep{Jenkins+etal:2010,Twicken+etal:2010}.

\citet{Lillo-Box+etal:2013} were able to exclude stellar oscillations as the source of the extra dimming by performing a zero-order cleaning of the solar-like oscillations of the star using a high-pass filter in Fourier space.  

\section{Methods}

For this analysis, we utilized the EXONEST software package \citep*{Knuth+etal:2012,Placek+etal:2013,Placek+etal:2014}, which employs a Bayesian inference engine capable of employing Nested Sampling \citep{Skilling:2006} and MultiNest \citep*{Feroz+etal:2009, Feroz+etal:2011, Feroz+etal:2013}, as well as Metropolis-Hastings Markov chain Monte Carlo (MCMC) \citep{Metropolis+etal:1953} and Simulated Annealing \citep{Otten:1989}.  The Multi-Nest engine was employed for this study for its ability to efficiently explore complicated parameter spaces.   

The EXONEST software package provides the ability to perform both Bayesian parameter estimation and model testing.  That is, given a photometric model that describes a hypothetical planetary system, EXONEST allows one to obtain model parameter estimates as well as an estimate of the Bayesian evidence, which is used to statistically test one model against another.  A model found to have an overwhelmingly large corresponding evidence value compared to the others is then said to be favored over the other model(s) describing the particular dataset \citep{Sivia&Skilling:2006, Knuth+etal:2015}.  The measure of one model's favorability over another is quantified by the Bayes' factor, which is the ratio of the model evidences.  Nested Sampling and Multi-Nest both compute log-evidences, so the Bayes' factor is calculated as the exponential of the difference between the log-evidences. 

EXONEST currently models four physical mechanisms in addition to transits and secondary eclipses that affect the photometric signal obtained from an exoplanetary system.  The first is the reflection of starlight off of the atmosphere or surface of the planet \citep{Seager+etal:2000, Jenkins&Doyle:2003, Seager:2010, Perryman:2011, Placek+etal:2014}, which is given by
\begin{equation}
\frac{F_R(t)}{F_\star} = \frac{A_g}{2} \frac{{R_p}^2}{r(t)^2} \left(1 + \cos \theta(t) \right) 
\end{equation}
 where $A_g$ is the geometric albedo, $R_p$ is the planetary radius, $r(t)$ is the star-planet separation distance, and $\theta(t)$ is the planetary phase angle.  The second is the thermally-emitted light from both the day- and night-sides of the planet \citep{Charbonneau:2005, Borucki+etal:2009, Placek+etal:2014}, given by
\begin{equation}
\frac{F_{D}(t)}{F_{\star}} = \frac{1}{2} \frac{B(T_d)}{B(T_{eff})}\left( \frac{R_p}{R_\star} \right)^2 \left( 1 + \cos\theta (t)\right)
\end{equation}
and
\begin{equation}
\frac{F_{N}(t)}{F_{\star}} =  \frac{1}{2} \frac{B(T_n)}{B(T_{eff})}\left( \frac{R_p}{R_\star} \right)^2 \left( 1 + \cos\theta (t)-\pi\right). 
\end{equation}
where $T_d$, and $T_n$ are the day- and night-side temperatures of the planet, $R_\star$ is the radius of the host star, and $B(T)$ is the observed thermal radiation in the \emph{Kepler} bandpass from a blackbody emitter at temperature $T$.  Both of these effects induce variations in the observed light as the planet orbits its host star and it goes through its phases (New, Crescent, Quarter, Full).  The third effect is Doppler beaming (or Boosting), which is a relativistic effect that causes an increase in observed flux when the host star is moving toward an observer, and a decrease when the star recedes from the observer \citep{Rybicki&Lightman:1979, Loeb&Gaudi:2003, Placek+etal:2014}.  This induces variations in the observed light curve since stars with planets around them orbit the common center of mass of the system, and thus periodically move toward and away from an observer.  The relative flux due to Doppler beaming is found by 
\begin{equation}
\frac{F_{B}(t)}{F_\star} =  1 + 4\beta_r(t)
\end{equation}
where $\beta_r$ is the stellar radial velocity along the observer's line of sight.  The fourth effect accounts for ellipsoidal variation, which are due to the gravitational tidal warping of the stellar surface due to the proximity of a massive planet \citep{Loeb&Gaudi:2003, Placek+etal:2014, Faigler&Mazeh:2011, Esteves+etal:2013,Shporer+etal:2011}.  A massive planet will warp the stellar surface into an ellipsoid whose semi-axis will follow the planet throughout its orbit.  This results in observed flux variations at $1/2$ the orbital period (twice the frequency) due to both the changing cross-sectional area of the star and an accompanying gravity darkening at the sub-planetary point and at the corresponding antipodal point.  This is approximated as
\begin{equation}
\frac{F_{E}(t)}{F_{\star}} =\alpha \frac{M_p}{M_\star} \left( \frac{R_\star}{r(t)} \right)^3 \sin^2i \cos2\theta(t) 
\end{equation}
where $M_p$ and $M_\star$ are the planetary and stellar masses, $i$ is the orbital inclination, and $\alpha$ is related to the linear limb-darkening coefficient $u$, and the gravity darkening coefficient $g$ by
\begin{equation}
\alpha = \frac{0.15(15+u)(1+g)}{3-u}.
\end{equation}

In addition to modeling the photometric effects in the case of both circular and eccentric orbits for \emph{Kepler}-91b alone, a two-planet model was created, which assumed that the orbit of a second body is of the same orbital period ($T$), eccentricity ($e$), inclination ($i$), and argument of periastron ($\omega$) as that of \emph{Kepler}-91b, but offset by an orbital phase of $\Delta \phi$.   

\emph{Kepler} data from quarters 1-16 were used to analyze \emph{Kepler}-91, which spans approximately $1550$ days, and includes 63,214 data points.  
The following stellar properties were also assumed: $u_1 = 0.65$, $u_2 = 0.05$, $u = 0.549$, $g = 0.733$ \citep{Claret&Bloeman:2011, Esteves+etal:2013}.  Here, the parameters $u$ and $g$ represent the linear limb-darkening coefficient and gravity-darkening coefficient respectively.  Similarly, the parameters $u_1$ and $u_2$ represent the quadratic limb-darkening coefficients.  The data were also folded on the accepted value of the orbital period, which assumes that both objects have an orbital period of $T = 6.24650$ days.

\subsection{Prior Probabilities and \\The Likelihood Function} \label{Priors&LogL}
 The Bayesian inference engine takes as inputs the prior probabilities for each model parameter and a likelihood function describing the expected noise distribution.

The prior probability distribution of a particular model parameter quantifies the knowledge one has about that parameter before analyzing the data.  For this study, each model parameter is assigned an uninformative uniform prior probability over a reasonable range as shown in Table \ref{Priors}.  Since stellar characteristics such as mass, radius, and effective temperature have all been estimated previously, we assign a Gaussian prior (Table \ref{Priors}) for these parameters defined by the published values in Table \ref{tbl:acceptedvalues}.

\begin{table}[h!]
\begin{center}
\begin{tabular}{l c}
  Parameter &   Distribution \\
\hline
Dayside Temp., $T_d$ (K)  	  &   $\mathcal{U}(0,6000)$ \\
Nightside Temp., $T_n$ (K) 	  &  $\mathcal{U}(0,6000)$ \\
Orbital Inclination, $\cos i$ 	  &  $\mathcal{U}(0,1)$ \\
Orbital Eccentricity, $e$		 &   $\mathcal{U}(0,1)$\\
Argument of Periastron, $\omega$ & $\mathcal{U}(0,2\pi)$\\
Initial Mean Anomaly, $M_0$    &   $\mathcal{U}(0,2\pi)$ \\
Stellar Mass, $M_\star$ $(M_\odot)$  & $\mathcal{N}(1.31, 0.10)$\\
Stellar Radius, $R_\star$ $(R_\odot)$ & $\mathcal{N}(6.30, 0.16)$\\
Planetary Radius, $R_p$ $(R_J)$ &  $\mathcal{U}(0,3)$ \\
Planetary Mass, $M_p $ $(M_J)$  &  $\mathcal{U}(0,10)$ \\
Geometric Albedo, $A_{g}$   &  $\mathcal{U}(0,1)$ \\
Standard Dev. of noise, $\sigma$ (ppm)  & $\mathcal{U}(0,1000)$ \\
Correlation Strength, $\epsilon$	&	$\mathcal{U}(-1,1)$\\
\hline
	\multicolumn{2}{c}{Fixed Values} 	\\
\hline 
Orbital Period, $T$	(d)	&	$6.24650$	\\
Quad. Limb Darkening $(u_1,u_2)$	&	$(0.65, 0.05)$\\
Linear Limb Darkening $u$ & 	$0.549$ \\
Gravity Darkening $g$ 	& $0.733$ \\
\hline
 \end{tabular}
\end{center}
\caption{Prior Distributions and fixed values for planetary, stellar and orbital parameters. The symbol $\mathcal{U}$ signifies a uniform distribution over the indicated range, while $\mathcal{N}$ represents a normal (Gaussian) distribution with mean and standard deviation inside the parentheses.} \label{Priors}
\end{table}

The likelihood function depends on the particular forward model and the expected nature of the noise.  In many situations one may assume that the noise is Gaussian distributed, however \emph{Kepler}-91 displays a significant amount of correlated (red) noise likely induced by stellar oscillations.  In order to accomodate the presence of correlated noise, we employ a nearest-neighbor approach introduced by \citet{Sivia&Skilling:2006} where the strength of correlations among nearest neighbors is described by the parameter $\epsilon$, which varies from $[-1,1]$.  Taking these nearest neighbor correlations into account one can obtain a log-likelihood function of the form \citep{Sivia&Skilling:2006}
\begin{multline}\label{logL_red}
\log L =  -\frac{N}{2}\log(2\pi\sigma^2) - \frac{(N-1)}{2}\log(1-\epsilon^2)  \\ + \frac{Q}{2(1-\epsilon^2)} 
\end{multline}
where $N$ is the number of data points, $\sigma^2$ the noise variance, and $Q$ is related to the sum of the squared residuals, $\chi^2$, by
\begin{equation}
Q = \chi^2 + \epsilon \left[ \epsilon(\chi^2 - \phi) - 2\psi \right].
\end{equation}
Here, $\phi$ is defined as the sum of the first and last squared residuals, and $\psi$ is the sum of the nearest neighbor residuals, which can be calculated using
\begin{equation}
\psi = \sum_{i = 1}^{N-1}R_i R_{i+1}.
\end{equation}
Note that when the correlation strength, $\epsilon = 0$, the log-likelihood function reduces to a Gaussian distribution of the form
\begin{equation}\label{logL_Gauss}
\log L = -\frac{N}{2}\log(2\pi\sigma^2) - \frac{\chi^2}{2\sigma^2}.
\end{equation}
For this study, both the correlated noise likelihood (\ref{logL_red}) and a Gaussian likelihood (\ref{logL_Gauss}) are used to analyze the Kepler-91b timeseries.

\begin{figure*}
\centering
\includegraphics[width=16cm]{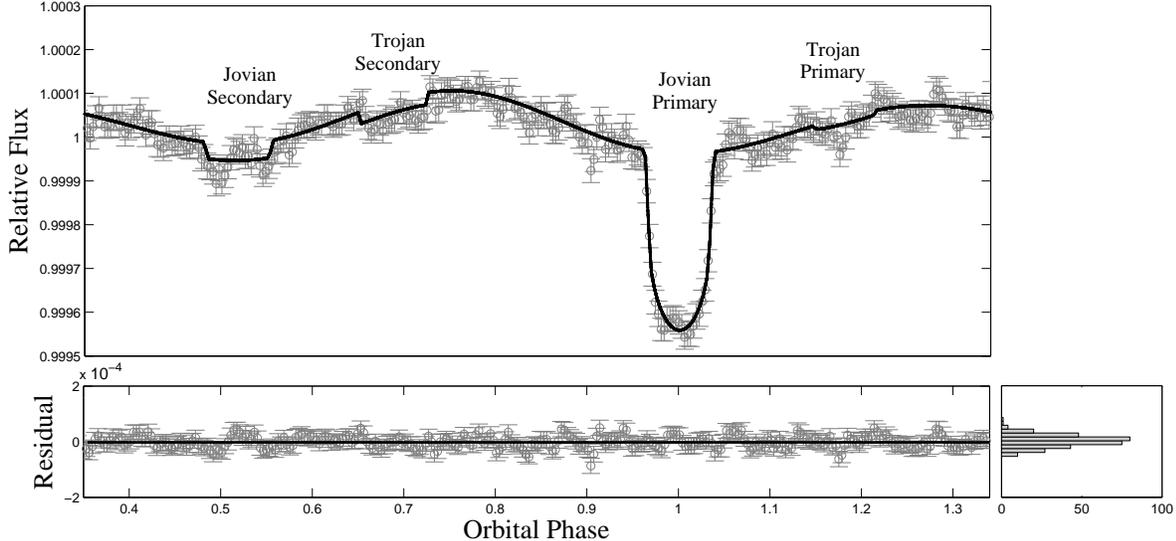}
\caption{\emph{Kepler}-91b plus Trojan fit for \emph{Kepler}-91. Dark grey data points represent the time series binned and averaged at the \emph{Kepler} cadence, and the black line is the model fit.  Notice that the third occultation, occurring at an orbital phase 0.7, has been modeled by EXONEST as a secondary eclipse as one would have expected from a Trojan.  Also, notice that the primary for the Trojan is much smaller in depth than the secondary, implying either a reduced transit depth due to the system's geometry or extremely hot day and night-sides. }
\label{fig:fit}
\end{figure*}

\section{Results}
This section summarizes our results from modeling the \emph{Kepler} photometry from quarters 1-16 and the published radial velocity measurements of \emph{Kepler}-91b \citep{Lillo-Box+etal:2014, Barclay+etal:2014}.  In addition, a short study of the stability of the hypothesized Trojan is presented using parameter values predicted by our analysis of the \emph{Kepler} data.

\subsection{From Photometry}\label{res_phot}

Results from our analyses are summarized in Table \ref{tbl:phot_parameters}.  The parameter estimates obtained from the one-planet model (\emph{Kepler}-91b only) indicate that \emph{Kepler}-91b is a hot-Jupiter with a mass of $M_p = 0.91 \pm 0.06 M_J$ and radius of $R_p = 1.39 \pm 0.02 R_J$ consistent with results from \citet{Lillo-Box+etal:2013}.  The dayside temperature was found to be $T_d = 2441.7 \pm 250.7K$. However, this may be significantly underestimated due to the mid-eclipse brightening event, which can be seen in Figure \ref{fig:fit} at orbital phase of $\sim 0.52$.  This day-side temperature is consistent with the equilibrium temperature calculated by
\citet{Lillo-Box+etal:2013}
assuming all incident thermal flux is re-radiated back into space ($T_{eq} = 2460^{+120}_{-40}K$). Based on the model log-evidences, the eccentric orbit model is significantly favored over the circular orbit model by a Bayes' factor of $\sim \exp(13)\thickapprox 440000$.  However, the derived orbital eccentricity was lower than that of \citet{Lillo-Box+etal:2013} at $e = 0.028 \pm 0.004$.

In the case of the two-planet model, which considers a photometric signal from both \emph{Kepler}-91b and the hypothesized Trojan, the log-evidence was such that the two-planet eccentric orbit model is favored by a Bayes' factor of $\sim \exp(16) \thickapprox 9$ million over the single planet eccentric orbit model.  The improvement in fitting is also demonstrated by the increased maximum likelihood of the solution and lower $\chi^2$ (see Table \ref{tbl:phot_parameters}). The predicted composite photometric signal from both objects is displayed in Figure \ref{fig:fit}.  Note that the third occultation that occurs around orbital phase 
$0.7$ 
is modeled by EXONEST as the secondary eclipse of the Trojan, as predicted, as well as a fourth occultation around an orbital phase of $1.2$, which represents the transit of the Trojan.  Parameter estimates associated with \emph{Kepler}-91b do not change significantly by adding another planet to the model.  The mass $M_p = 0.99 \pm 0.07M_J$, radius $R_p = 1.38 \pm 0.02R_J$, albedo $A_g = 0.39 \pm 0.17$, and dayside temperature $T_d = 2513.2 \pm 317.9K$ are all within $1\sigma$ of the estimates from the one-planet model.  Parameter estimates from the two-planet eccentric model suggest that the hypothesized Trojan has a mass of $M_{p} = 0.025 \pm 0.019M_J$ or $M_{p} = 8.89 \pm 6.04M_\oplus$, and radius of $R_{p} = 0.28 \pm 0.05R_\oplus$ giving it a mean density of $\bar{\rho} = 1.98 \pm 0.77 g/cm^3$, which implies that, if this is a planet, it could potentially be classified as sub-Neptunian. This estimate of the density is consistent with results from 
\citet{Rogers:2014}, 
who showed using the known populations of sub-Neptune planets that planets with radii greater than 
$1.6R_\oplus$
are most likely not rocky, but gaseous. 
In addition, the presence of a Trojan body is not only supported by the log-evidence calculations, but also by the model prediction for the planet-Trojan phase difference, which is $\Delta \phi = 65.4 \pm 1.7$ degrees.  

This model also predicted very high day and night-side brightness temperatures of $T_d = 5184.6 \pm 531.5$K and $T_n = 2372.6 \pm 1283.6$K, respectively, which will be further discussed in Section \ref{sec:disc}.  As for the primary planet, the two-planet model predicts \emph{Kepler}-91b to have day and night-side temperatures that are within $1\sigma$ of each other ($T_d = 2513.2 \pm 317.9K$ and $T_n = 2871.8 \pm 183.6$), which is not unexpected for short-period hot Jupiters around red giant stars \citep{Spiegel&Madhusudhan:2012}.  

This model also predicted an unexpectedly high albedo for the Trojan $A_g = 0.49 \pm 0.28$, which is likely due to the strong degeneracy between the albedo $A_g$ and the day-side temperature $T_d$.  As indicated by the large uncertainties, there is simply not enough information in the data for EXONEST to disentangle the reflected and emitted components and to well-determine the geometric albedo of the hypothetical Trojan. 

\begin{figure*}
\includegraphics[width=17cm]{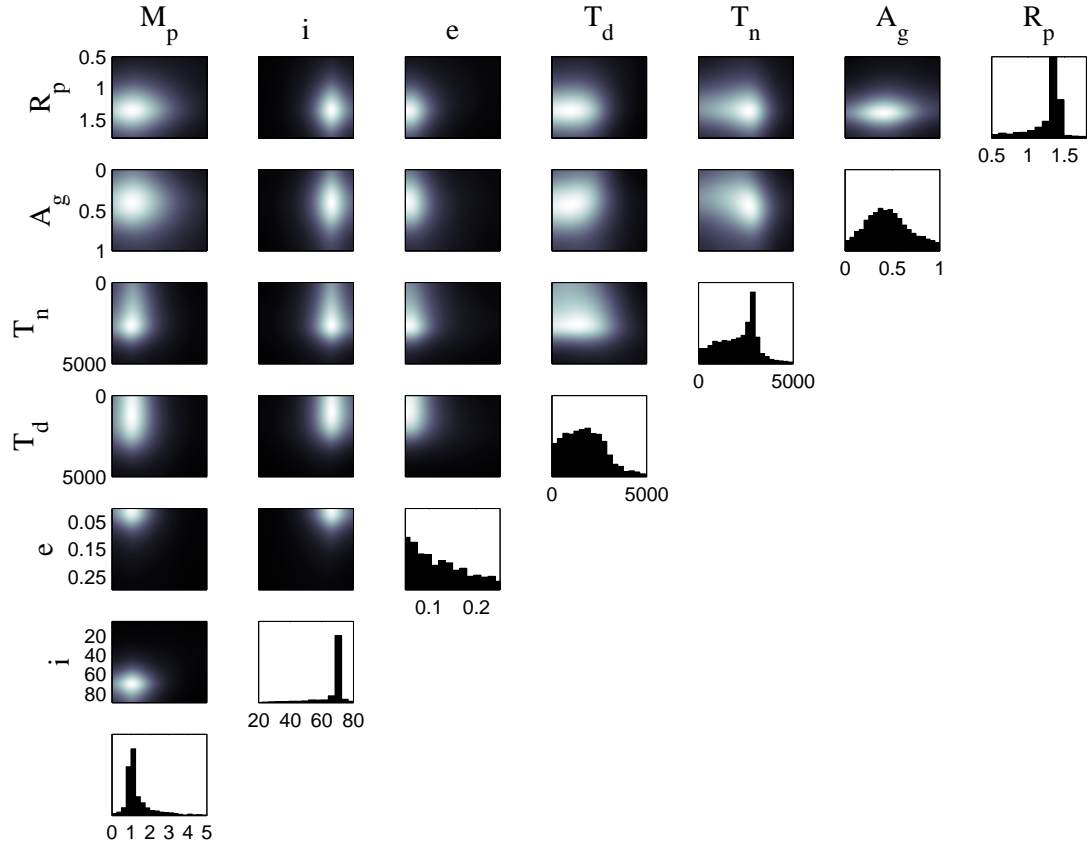}
\caption{Two-dimensional representations of the posterior parameter estimates associated with \emph{Kepler}-91b in the case of the two-planet model.  Of note, is the poorly-constrained geometric albedo, which is caused by the degeneracy between the thermal emission and reflected light effects. This degeneracy is further illustrated in the $T_d$ vs. $A_g$ plot where there is a slight banana-shaped ridge and causes large error bars on their respective parameter estimates. }
\label{Jovian}
\end{figure*}

\begin{figure*}
\includegraphics[width=17cm]{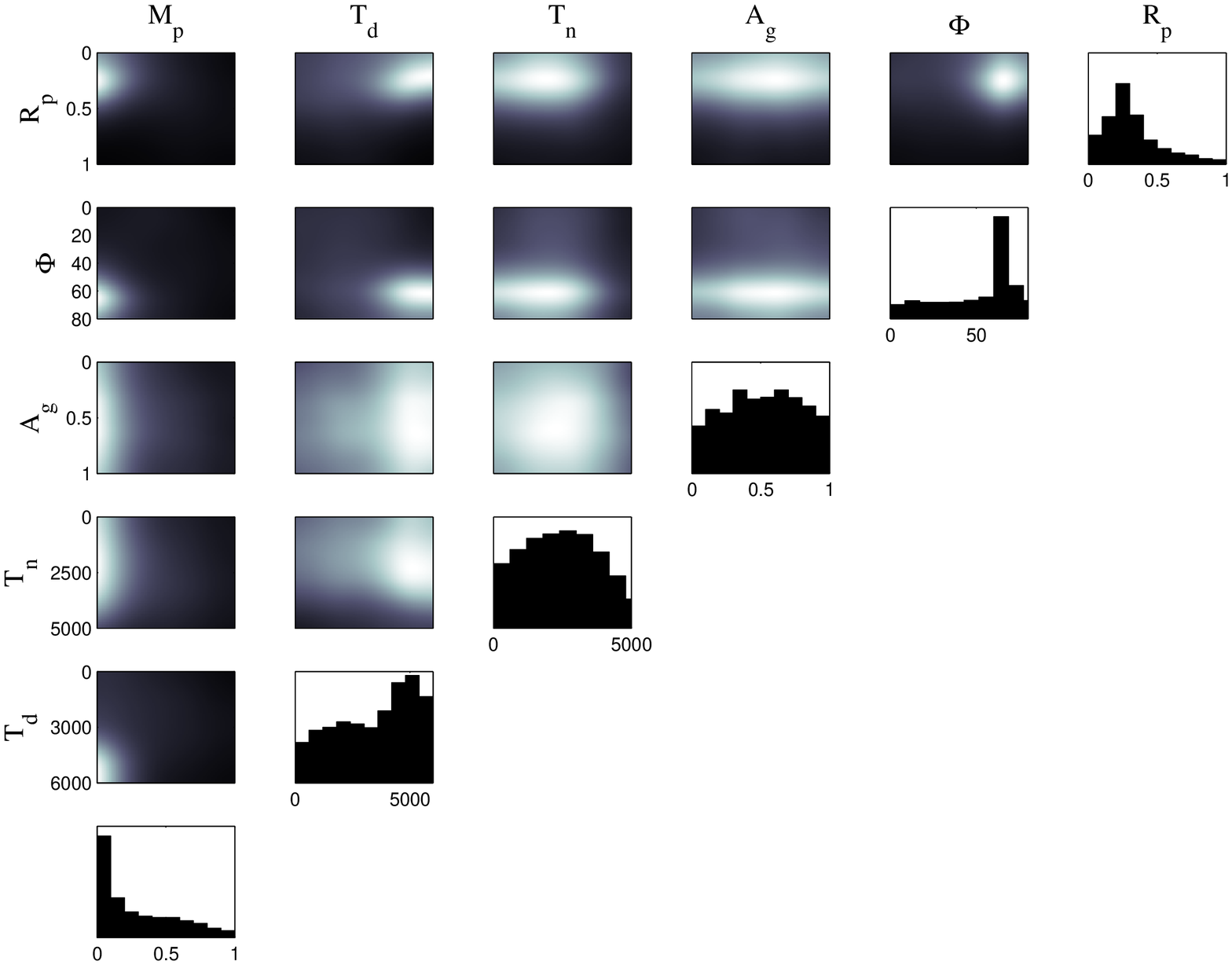}
\caption{Two-dimensional representations of the posterior parameter estimates associated with the hypothetical Trojan in the case of the two-planet model.  Notice the strong peak in the phase angle of the Trojan at $\approx 60^o$. Also, the first column illustrates that the mass of the Trojan must be very small, which is likely due to the apparent lack of ellipsoidal variations detected in the phase-folded light curve.   }
\label{Trojan}
\end{figure*}

Two-dimensional representations of the posterior probability of the model parameters are illustrated in Figures \ref{Jovian} and \ref{Trojan}.   Notice the degeneracy between $T_d$ and $A_g$ in Figure \ref{Jovian}.  A planet can have a higher day-side temperature and lower geometric albedo and output the same flux as a highly reflective planet (high geometric albedo) with a low day-side temperature.  One can also see that many of the parameters associated with the Trojan exhibit broad ridge-like structures, which explain our inability to precisely estimate them.

Based on the estimate of the Jovian mass obtained from the two-planet model, the expected long- and short-period librations of the Trojan can be calculated to be $T_{\rm long} = 90.32 \pm 4.80$ days, and $T_{\rm short} = 6.254 \pm 0.001$ days, respectively.  The short period epicyclic librations have approximately the same period as the orbital period of the two planets around the host star.  The long period librations were searched for using a Lomb-Scargle periodogram, however no pronounced periodicities were discovered.


The presence of correlated noise was verified by the algorithm, which predicted a nearest-neighbor correlation strength, $\epsilon$, of approximately $0.365 \pm 0.004$ for all models.  It is certainly possible that the Jovian has set up oscillations in the host star.  However, for these oscillations to manifest themselves as a pulse wave with a pulsewidth $\tau$ that mimics the Trojan's secondary eclipse, the star would need to be oscillating according to something of the form
$$f(t) = C_1 + C_2 \sum_{n=1}^\infty \frac{2}{n \pi} \sin( \frac{n \pi \tau}{T} ) \cos( \frac{2 n \pi }{T} t ) $$
where $T$ is the period of the Jovian and $C_1$ and $C_2$ are constants.  It is highly unlikely that some resonance excited by a driving oscillation at period $T$ would result in an overall stellar oscillation consisting of frequency components with amplitudes precisely modulated by a quantity $\tau$ (appearing in the functional form above) that is in agreement with the transit/eclipse duration of the Jovian, which relies on the viewpoint-dependent impact parameter.

A Gaussian likelihood that neglects correlations among the observed data was also used in the analysis. With the Gaussian likelihood, the Jovian+Trojan model was again favored by a Bayes' factor of $\sim \exp(8)\thickapprox 3000$ over the other two photometric models.  The parameter values associated with the Jovian and Trojan were also similar, not deviating more than $1\sigma$ from the estimates obtained using the correlated noise likelihood function (in Table \ref{tbl:phot_parameters}).  Comparing log-evidences, the correlated noise likelihood ($\log Z = 411\,937.4 \pm 0.9$) is favored over the Gaussian ($\log Z = 407\,408 \pm 0.8$) by a Bayes' factor of $\sim \exp(4\,500)$, indicating that the correlated noise model provides a significantly better description of the observed data. 

\begin{table*}
\centering

\begin{tabular}{l c c c  c  }
					&  \multicolumn{2}{c}{\textbf{Two-Planet Model}}					 &\begin{tabular}{@{}c@{}} \textbf{One-Planet Model} \\  \textbf{Eccentric} \end{tabular}	&  \begin{tabular}{@{}c@{}} \textbf{One-Planet Model} \\  \textbf{Circular} \end{tabular}  \\
\hline

Parameter 			&\emph{ Kepler-91b} 			& \emph{Trojan Candidate}   		&\emph{Kepler-91b} 					& \emph{Kepler-91b}  \\
\hline
$i$	($deg$)			&$69.81 \pm 0.18$				& ... 					&$69.80 \pm 0.14 $					&$69.70^{+0.14}_{-0.12}$  \\
$\omega$ ($rad$)		&$3.71 \pm 0.29$				& ... 					&$3.03 \pm 0.27$						&$...$  \\
$e$					&$0.028 \pm 0.004$			& ... 					&$0.029 \pm 0.004$					&$...$  \\
$M_p$ ($M_J$)			&$0.99 \pm 0.07$				& $0.025 \pm 0.019$	&$0.91 \pm 0.06$						&$0.93 \pm 0.06$  \\
$R_p$ ($R_J$)			&$1.38 \pm 0.02$				&$0.26 \pm 0.06$		&$1.39 \pm 0.02$						&$1.39 \pm 0.02$  \\
$T_d$ ($K$)			&$2513.2 \pm 317.9$			&$5184.6 \pm 531.5$ 		&$2441.7 \pm 250.7$				&$2433.4 \pm 249.7$  \\
$T_n$ ($K$)			&$2871.8 \pm 183.6$			&$2372.6 \pm 1283.6$		&$2348.2 \pm 684.2$				&$1638.3 \pm 887.7$  \\
$A_g$ 				&$0.39 \pm 0.17$				&$0.49 \pm 0.28$		&$0.39 \pm 0.15$						&$0.22 \pm 0.11$  \\
$\Delta \phi$	($rad$)	&$...$						&$65.4 \pm 1.7$		&$...$								&$...$  \\
\hline
$\log Z$	&  	\multicolumn{2}{c}{$411937.4 \pm 0.9$}						 & $411921.3 \pm 0.9$					&$411908.6 \pm 0.9$  \\
$\chi^2$ &	\multicolumn{2}{c}{$0.0092$}									& $0.0092$							&$0.0092$	\\
$\log L_{max}$ &	\multicolumn{2}{c}{$4.1199e05$}							&$4.1197e05$							& $4.1195e05$	\\
$\epsilon$ 	& \multicolumn{2}{c}{$0.365 \pm 0.003$}						&$0.365 \pm 0.004$					&$0.365 \pm 0.004$  \\
\end{tabular}
\caption{Model parameters and log-evidences ($\log Z$) for all three models applied to the \emph{Kepler}-91 light curve. In addition to parameter estimates, $\chi^2$, which is the sum of the squared residuals, and $\log L_{max}$, which represents the maximum of the likelihood function --- the probability of observing a dataset given a model set of model parameter values.  The lower the $\chi^2$ value, the better the fit whereas the higher the $\log L$ the better the fit.  For this particular set of photometric models, the two-planet model has the largest $\log Z$ making it the most favored model from a Bayesian model selection standpoint. The two-planet model also has the lowest $\chi^2$, and highest $\log L_{max}$ values making it a better fit compared to the two one-planet models. The nearest-neighbor correlated noise log-likelihood function described in Section \ref{Priors&LogL} was used for each of these simulations and the correlation strength $\epsilon$ was consistenly estimated to be $0.365 \pm 0.004$ for each model. }
\label{tbl:phot_parameters}
\end{table*}

\subsection{From Radial Velocity}\label{res_rv}
Bayesian model testing was also performed on the published radial velocity measurements \citep{Lillo-Box+etal:2014, Barclay+etal:2014} in order to determine if a model that includes the radial velocity signal from a Trojan planet would be favored over a model that neglects a Trojan.  Assumed quantities for these simulations include only the stellar mass ($M_\star =1.31 \pm 0.10 M_\odot$),  and orbital period ($T = 6.24650$ d) as opposed to the analysis performed by \citet{Lillo-Box+etal:2014}, which assumed the eccentricity to be that which was estimated from their analysis of photometry.  Based on model log-evidences, the eccentric one-planet model was slightly favored over the circular model by a Bayes' factor of $\sim \exp(1.5) \approx 4.5$ in agreement with the photometric results.  The eccentric one-planet model predicts an eccentricity of $e = 0.041 \pm 0.028$, which is also in agreement (within uncertainty) with results presented in the previous section as well as the photometric analysis performed by \citet{Lillo-Box+etal:2013}.  This model also predicts a minimum planetary mass associated with \emph{Kepler}-91b of $M_p \sin i = 0.86 \pm 0.09 M_J$, which along with the orbital inclination obtained from photometry, implies a true mass of $M_p = 0.92 \pm 0.10 M_J$, which is in agreement with the results from photometry.
These two models were then tested against an eccentric two-planet model that assumes the second planet to be in the same orbit as the Jovian, but trailing it by a phase angle $\Delta \phi$.  This model was neither favored over nor rejected by the eccentric one-planet model.  However, it is interesting that adding an additional two parameters to describe the hypothetical Trojan did not significantly change the log-evidence.  The one-planet eccentric model yielded a higher log-evidence, with a difference in log-evidence between the two models being $\sim \exp(1.53)$.   It should also be noted that attempting to fit two radial velocity signals with the same period will result in significant degeneracies as can be seen in the semi-amplitudes estimated in Table \ref{tbl:rv_parameters}.  The predicted semi-amplitudes correspond to uncertainties in $M_p \sin i$ that are approximately $50\%$ of the mean value.  However, this does not affect ones ability to perform Bayesian model testing.  The two planet model also predicted a phase difference of $\Delta \phi = 79.06^o \pm 37.8^o$.  While the uncertainty here is large, this doesn't rule out the expected phase difference for a Trojan of $60^o$.  These radial velocity measurements neither confirm or reject the Trojan hypothesis.  
In addition to comparing log-evidences, another common way to compare different models is to look at the $\chi^2$ value, which represents the sum of the squared residuals.  Having the lowest $\chi^2$, the two-planet model represents a better fit to the observed data.  In a similar fashion, one may compare the log-likelihoods of a set of models, which describes the probability of observing a particular dataset given a set of model parameters.  The higher the log-likelihood, the better the fit to the data.  The two-planet model clearly has the highest maximum log-likelihood, but is not favored in log-evidence because the fit was not improved sufficiently to overcome the penalty of adding additional parameters to the model, which is taken into account when calculating the log-evidence.

\begin{figure*}
\centering
\includegraphics[width=12cm]{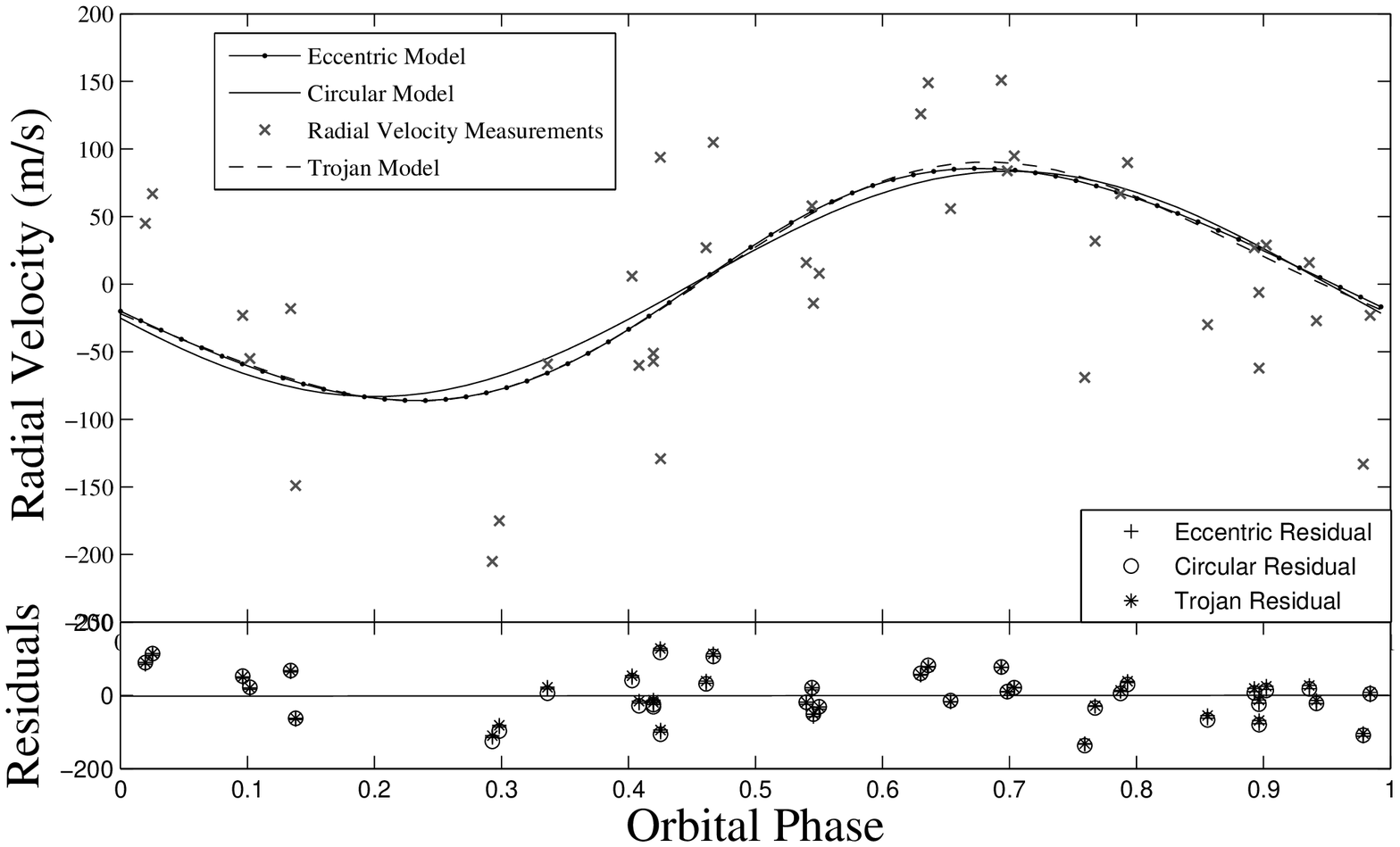}
\caption{Model fits to the radial velocity measurements of \emph{Kepler}-91 \citep{Lillo-Box+etal:2014}.  The three models displayed are the one-planet in eccentric (solid line with dots; $\log Z = -240.31 \pm 0.35$, $\chi^2 = 1.5769 \times 10^5$, $\log L_{max} = -225.6953$) and circular (solid line; $\log Z = -241.50 \pm 0.36$, $\chi^2 = 1.6622 \times 10^5$, $\log L_{max} = -226.6576$) orbits, as well as the two-planet eccentric model (dashed; $\log Z = -241.53 \pm 0.26$, $\chi^2 = 1.5732 \times 10^5$, $\log L_{max} = -224.6390$) that attempts to describe the observed radial velocity measurements by including the Jovian and a candidate Trojan planet. The bottom window of the plot displays the corresponding residuals for each model. }
\label{fig:fit}
\end{figure*}

\begin{table*}
\centering
\begin{tabular}{l c c c  c}
					&  \multicolumn{2}{c}{\textbf{Two-Planet Model}}					 &\begin{tabular}{@{}c@{}} \textbf{One-Planet Model} \\  \textbf{Eccentric} \end{tabular}	&  \begin{tabular}{@{}c@{}} \textbf{One-Planet Model} \\  \textbf{Circular} \end{tabular} \\
\hline

Parameter 			&\emph{ Kepler-91b} 			& \emph{Trojan Candidate}   &\emph{Kepler-91b} 				& \emph{Kepler-91b}\\
\hline
$K$	($m/s$)			&$50.94 \pm 27.98$			&$58.77 \pm 26.56	$	&$83.67 \pm 9.77 $					&$82.28 \pm 11.57$\\
$\omega$ ($rad$)		&$3.02 \pm 1.69$				& ... 					&$2.82 \pm 2.23$						&$...$\\
$e$					&$0.047 \pm 0.028$			& ... 					&$0.041 \pm 0.028$					&$...$\\
$M_p \sin i$ ($M_J$)	&$0.48 \pm 0.23$				& $0.61 \pm0.28 $			&$ 0.86 \pm 0.09 $						&$0.86 \pm 0.12 $\\
$\Delta \phi$	($deg$)	&$...$			 			&$79.06 \pm 37.8$		&$...$								&$...$\\
$\gamma$ ($m/s$)		&$-31.00 \pm 0.04$			& $...$				&$-62.010 \pm 0.006$					&$-62.011 \pm 0.008$\\
\hline
$\log Z$	&  	\multicolumn{2}{c}{$-241.53 \pm 0.26$} 						& $-240.31 \pm 0.35$					&$-241.50 \pm 0.36$\\
$\chi^2$ &	\multicolumn{2}{c}{$1.5732 \times 10^5$}						& $1.5769 \times 10^5$				&$1.6622 \times 10^5$				\\
$\log L_{max}$ &	\multicolumn{2}{c}{$-224.6390$}								&$-225.6953$							& $-226.6576$			\\
\end{tabular}
\caption{Model parameters for all three models applied to the \emph{Kepler}-91 radial velocity measurements.  In addition to parameter estimates, included is the $\chi^2$ value, which is the sum of the squared residuals, and $\log L_{max}$ value, which represents the maximum of the likelihood function --- the probability of observing a dataset given a model set of model parameter values.  The lower the $\chi^2$ value, the better the fit whereas the higher the $\log L$ the better the fit. For this particular set of radial velocity models, the Eccentric one-planet model has the largest $\log Z$ making it the most favored one-planet model from a Bayesian model selection standpoint. Notice that the two-planet model also has the lowest $\chi^2$, and highest $\log L_{max}$ values making it a better fit compared to the two one-planet models. }
\label{tbl:rv_parameters}
\end{table*}

\subsection{Stability of Trojan Configuration}\label{Trojan Stability Section}
In order to test the Trojan planet hypothesis further, a preliminary stability study was performed.  One hundred ($100$) orbital configurations were generated by sampling the posterior obtained from the \emph{Kepler} photometry.  Initial conditions for each orbital configuration were determined by sampling from the posterior distributions obtained in the photometric two-planet simulations for $M_\star$, $R_\star$, $M_{\rm jov}$, $M_{\rm tro}$, $e$, $a/R_\star$, and $\Delta \phi$ with the means and standard deviations being those that are listed in Table \ref{tbl:phot_parameters} for those that correspond to the photometric model parameters, and Table \ref{tbl:acceptedvalues} for those which are previously published values.  Each orbital configuration was tested for stability using the following cuts. The semi-major axes of the Jovian and hypothetical Trojan orbits cannot increase or decrease by more than 30\% of the original value, the eccentricity of the orbits for either object cannot exceed one ($e\ngeq 1$), which would indicate a parabolic or hyperbolic trajectory, and the maximum allowed change in energy of the system was $10^{-4}$ \citep{Barnes&Quinn:2004}.  The equations of motion for the three-body problem were integrated up to 50,000 days, which corresponds to approximately $8,000$ orbits of the Jovian and Trojan.  All orbital configurations for the Trojan were found to be stable over this time period.  Furthermore, the results predict librational periods of $T_{l\rm ong} = 80.36 \pm 11.21$ days and $T_{\rm short} = 5.71 \pm 0.83$ days, which both agree with the theoretical values predicted by equations (\ref{long libration}) and (\ref{short libration}).  The relatively large uncertainties on these librational periods are likely due to the sampling of the initial conditions from the posterior.  In the future longer integration times will be necessary to solidify the stability of such an orbital configuration.  

\subsection{Stability of Exomoon Configuration}\label{Exomoon Stability Section}
A similar stability study was performed to investigate the exomoon hypothesis. Each parameter was again sampled from the posterior obtained from the photometric simulations. This excludes the mass and semi-major axis of the hypothesized exomoon since they were not modeled in the light curve.  Due to the fact that the occultations associated with the hypothesized third body are observed in the light curve folded on the period of the Jovian, it would be reasonable to assume that the two bodies are in resonance.  Therefore, the period of the exomoon was assumed to be $T_m = n T_J$, where $n$ ranges from $n = [0.1,3]$ in increments of $0.1$ and $T_J = 6.2465$ days. For each value of $n$, $100$ orbital configurations were generated. 

For an exomoon with mass $M_m = 0.1 M_J$, there were zero stable orbits for each value of $n$.  In the case of the 1:10 resonance, every configuration resulted in the moon crashing into the Jovian. For the 1:5 resonance, $78\%$ of the configurations resulted in the moon crashing into the Jovian, and $22\%$ resulted in the moon being expelled into a more distant orbit around the host star. The remaining configurations all resulted in the moon being shot out to more distant orbits around the host star, likely due to the fact that it's semi-major axis was larger than the radius of the Hill sphere of the Jovian, which is approximated by
\begin{equation}
r_H \approx a(1-e)\sqrt[3]{\frac{M_{\text{jov}}}{3M_\star}}.
\end{equation}
A moon that is outside of the Hill sphere of a planet, is not gravitationally bound to that planet.  Based on the posterior estimates from photometry, the radius of the Hill sphere in the case of \emph{Kepler}-91b is $r_H = 0.0041 \pm 0.0004$ AU.  This would translate to orbital periods for the moon of $T_m = 0.51 \pm 0.08 T_{\text{jov}}$, which verifies the unstable behavior observed in the numerical simulations.

\begin{table*}[t!]
\begin{center}
\begin{tabular}{l c c c c c }
  Subset &   $logZ_{\text{two}} - logZ_{\text{one}}$  & $\chi^2_{\text{two}}$  (ppm)& $\chi^2_{\text{one}}$ (ppm) & $\log L_{max, two}$	& $\log L_{max, one}$ \\
\hline
Even - All Data & +11.5 	& $\bf{4559.10}$	& 4562.24	&$\bf{2.04710e05}$   &$2.04693e05$	\\
Odd - All Data &  - 1.5 	&$4634.87$	& \bf{4634.59}		&$\bf{2.07267e05}$ &$2.07258e05$\\
Even - First Half & +13.9 & $\bf{2289.05}$	& $2293.96$	&$\bf{1.02945e05}$ &$1.02924e05$\\
Odd - First Half & -6.9 	&  $2339.96$	& $\bf{2339.54}$	&$\bf{1.02903e05}$ &$1.02905e05$\\
Even - Second Half & +3.2	&$\bf{2304.50}$& $2307.24$ &$\bf{1.04567e05}$ &$1.04557e05$\\
Odd - Second Half & -0.6	&$2242.90$	& $\bf{2242.19}$ &$\bf{1.01635e05}$ &$1.01633e05$\\
\hline
 \end{tabular}
\end{center}
\caption{Model testing results on odd/even subsets of the 16 quarters of \emph{Kepler} observations. Positive values indicate that the two-planet model was favored over the one-planet model. The two-planet model is favored in the even periods, and the odd periods favor the one-planet model. Also listed are the $\chi^2$ values for each simulation. Notice the odd/even differences are evident in both log-evidence and $\chi^2$ values.   } \label{tbl:even_odd}
\end{table*}

\subsection{Odd-Even Effects \\ and Systematic Noise}\label{sec:even_odd}
As shown in Figure \ref{fig:half periods}, there are odd/even effects that appear in the light curve folded at twice the orbital period of the Jovian. These appear to occur halfway through the transit of the Jovian and apparent secondary of the Trojan (see Figure \ref{fig:half periods}) and when binned at the \emph{Kepler} cadence and folded on the accepted period of the Jovian apparently disappear. To test whether or not these are significant in affecting the model testing results, we tested the one-planet model against the two-planet model using six different subsets of the entire time-series.  The subsets were defined as follows: the even/odd periods of the Jovian over the all quarters, the even/odd periods of the Jovian over the first eight quarters, and the even/odd periods of the Jovian over the last eight quarters for a total of six subsets. Results are displayed in Table \ref{tbl:even_odd}. 

Positive differences in the log-evidences indicate that the two-planet model is favored over the one-planet, whereas negative differences indicate that the one-planet model is favored. Clearly the odd/even effects are influencing the model testing results, since the two-planet model is favored over the one-planet model in the even periods, but not in the odd periods. This is also apparent in the model fits as the minimum chi-squared values (bolded in Table \ref{tbl:even_odd}) alternate between even and odd periods.  Since the libration period is estimated to be on the order of the orbital period of the Jovian, it is unlikely that the libration of a Trojan companion could explain this result. It is very possible that there is an unmodeled stellar effect or star-planet interaction that is responsible for these odd/even differences. The maximum log-likelihoods for each model are also listed in Table \ref{tbl:even_odd}. Unlike the chi-squared values, these do not alternate between odd and even periods likely due to the fact that there are free parameters affecting the likelihood function. Namely the standard deviation of the noise, $\sigma$, and the correlation coefficient, $\epsilon$.  The one-planet model has fewer degrees of freedom compared to the two-planet model. As such there may be features in the data that the model cannot fit, and thus treats as noise by altering the estimated parameters that correspond to the likelihood function. 

Since there is suspected to be a significant amount of systematic noise in the light curve of \emph{Kepler}-91b, these extra dimming events may be caused by some unknown systematic effect(s). If the signal in the phase-folded light curve is truly from a Trojan companion, one should only rarely see similar dimming events (in depth and duration), when the light curve is folded on other random periods. 

\begin{figure*}[h!]
\begin{center}$
\begin{array}{cc}
\multicolumn{2}{c}{\includegraphics[width = 9cm]{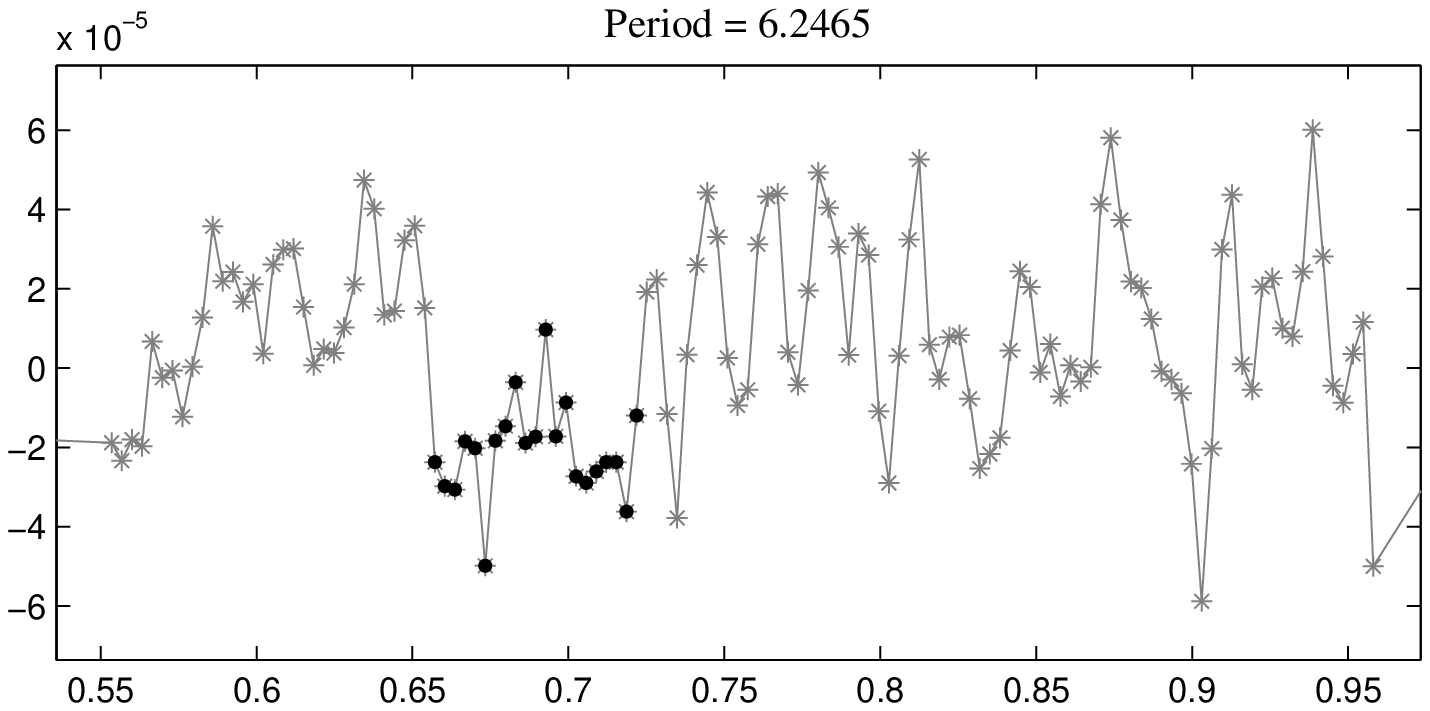}} \\
\includegraphics[width = 8cm]{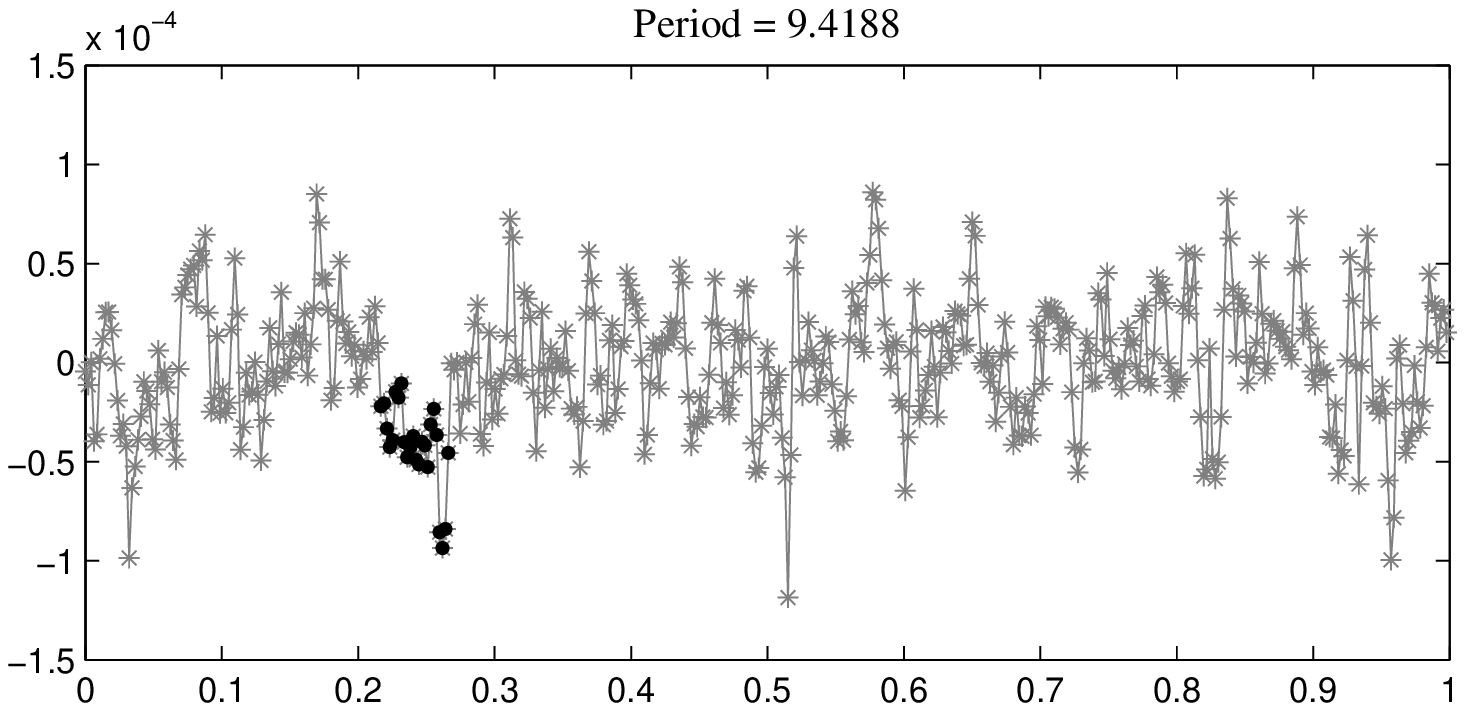} & \includegraphics[width = 8.0cm]{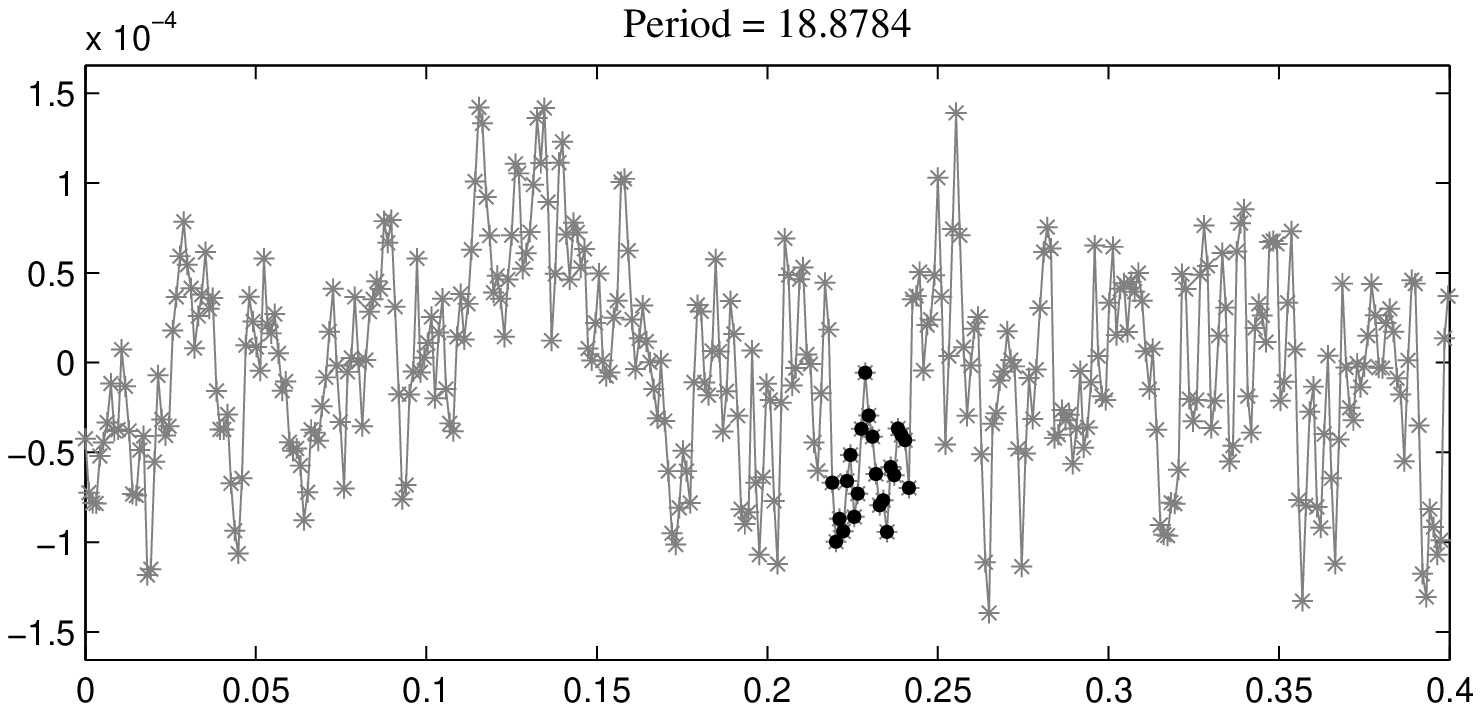}\\
\includegraphics[width = 8.0cm]{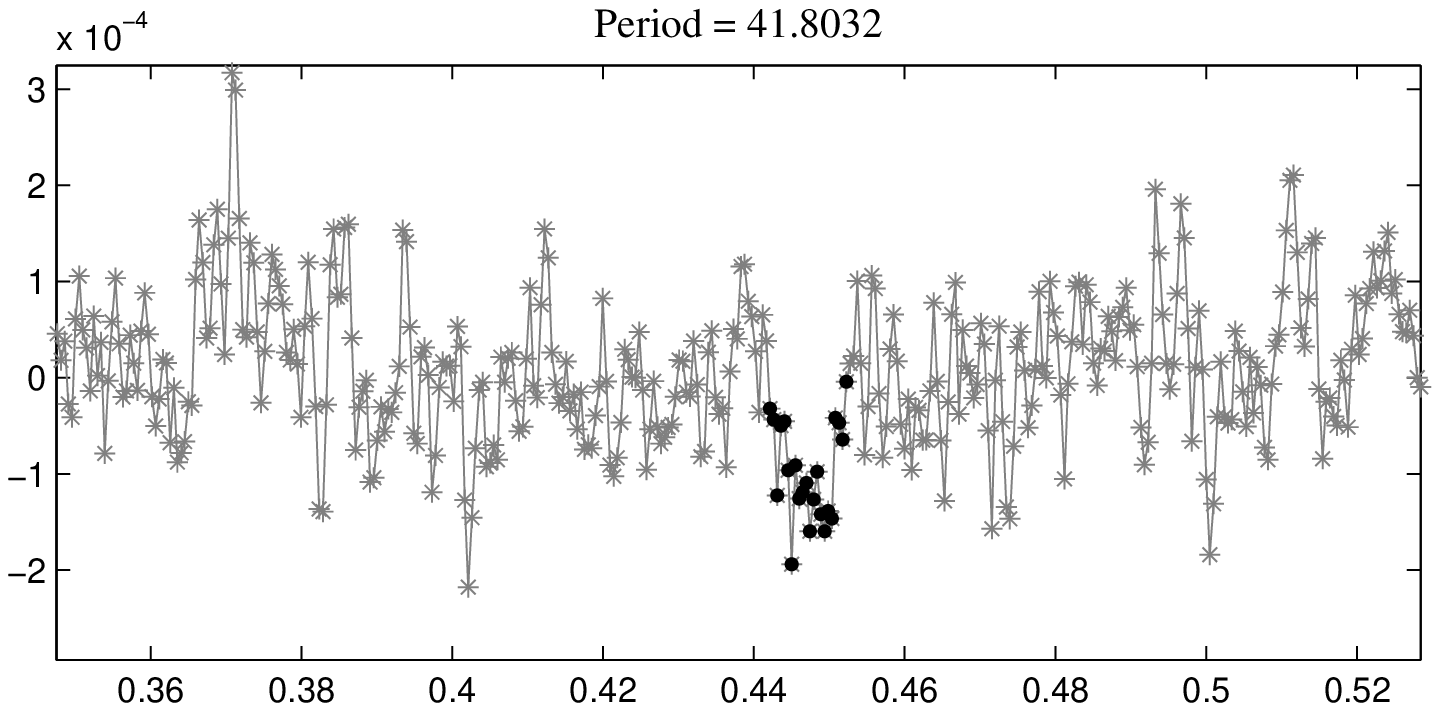} & \includegraphics[width = 8.0cm]{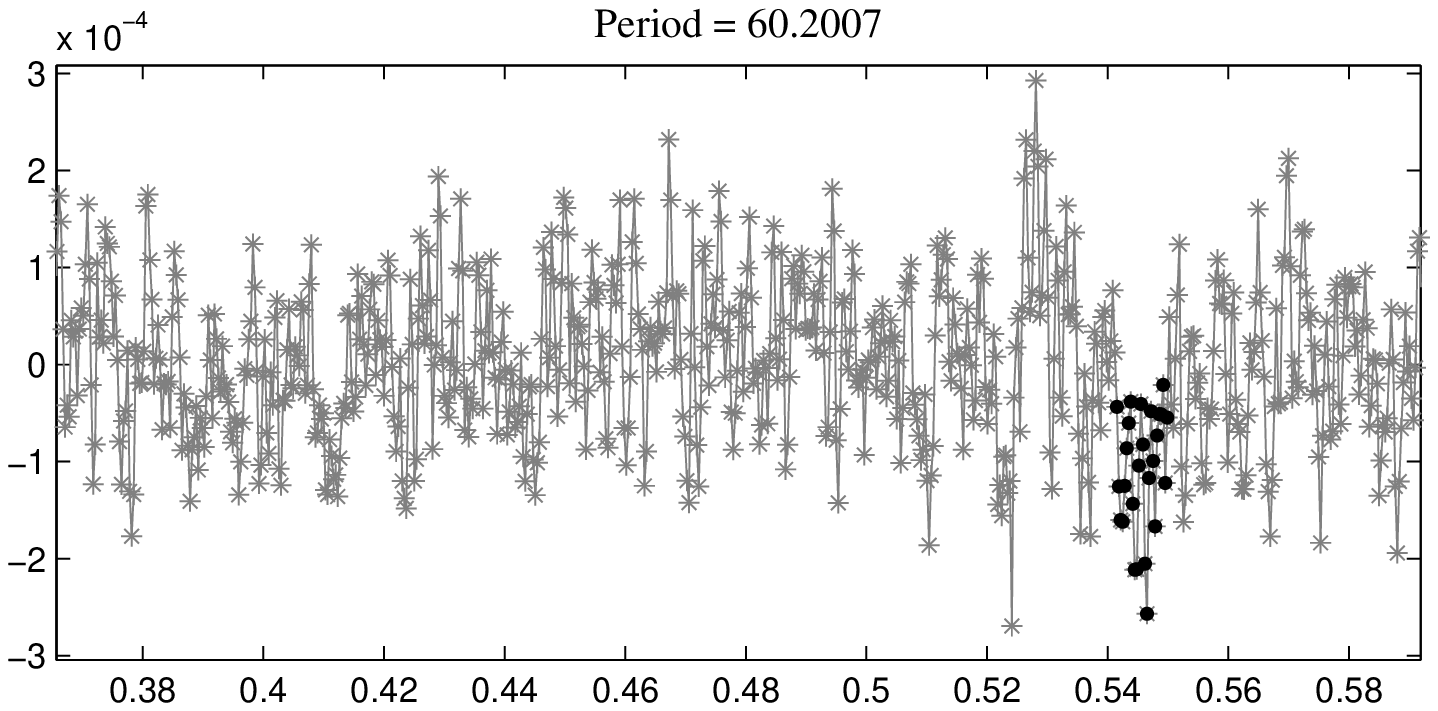} \\ \includegraphics[width = 8.0cm]{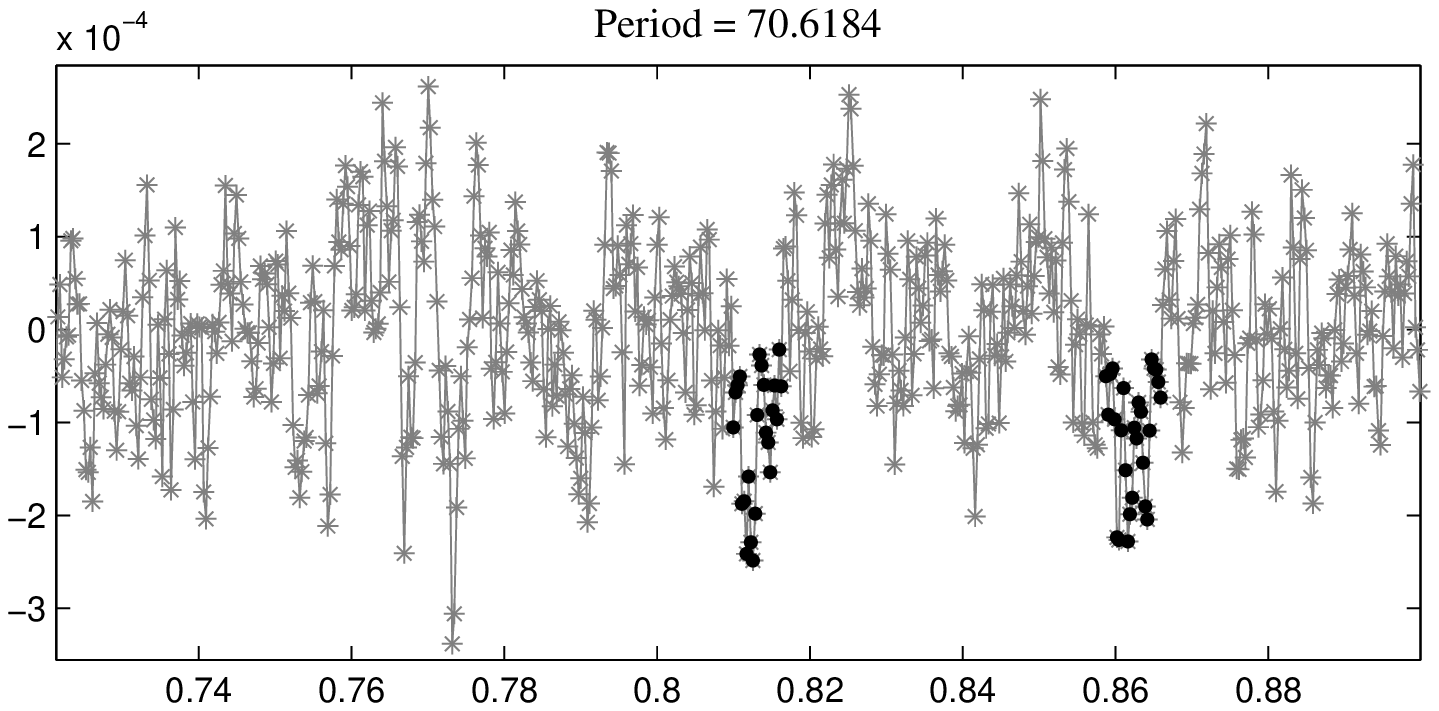} & \includegraphics[width = 8.0cm]{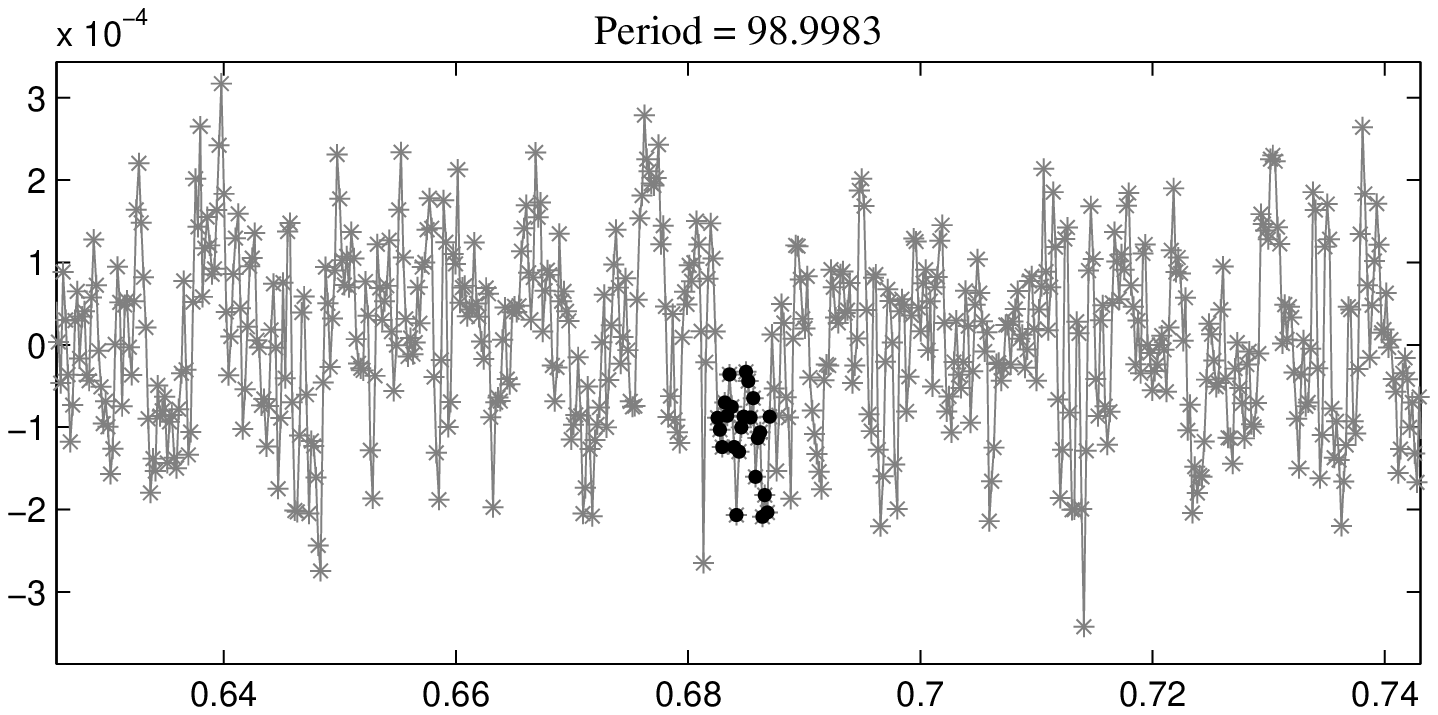} \\
\end{array}$
\end{center}
\caption{Six examples of the randomized folding routine including a zoomed in view of the apparent Trojan signal (top). The grey data points represent the \emph{Kepler}-91b residuals folded on the period listed above each window binned at the \emph{Kepler} cadence. Detected dimming events are labeled by black dots. These events are all approximately the same duration (0.495 days), however the durations appear different in each window as they are folded at different periods. For longer periods, the duration will appear to decrease. } \label{fig:dimming_events}
\end{figure*}

\begin{figure}[h!]
\centering
\includegraphics[width=7cm]{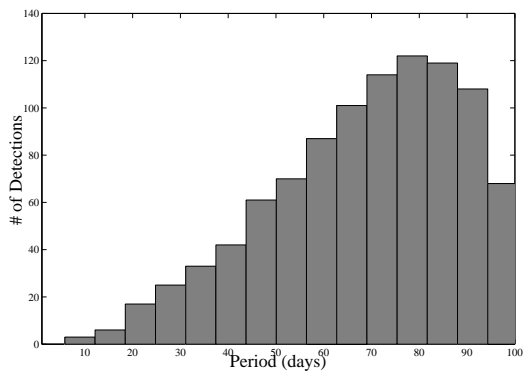}
\caption{Histogram of the detected dimming events in the light curve of \emph{Kepler}-91b folded at random periods. Notice that the events are more abundant at periods of approximately 70 days and very rare near the orbital period of the Jovian (6.2465 days). }
\label{hist_detections}
\end{figure}

A transit detection routine was created to search for such eclipses and determine their frequency at various periods. First, the signal from the Jovian was subtracted from the light curve using the best-fit model from the photometric analysis in Section \ref{res_phot}. Then, the residuals corresponding to the intervals over which the Jovian was in transit and when it was in secondary eclipse were removed. These residuals were then folded on random periods ranging from [0,100] days. Detections were deemed to have occured if the data points dipped below the mean of the entire dataset, at which time the area under the curve would be computed. A larger transit would correspond to a smaller, more negative, area under the curve. If any detections corresponded to a greater area under the curve than the apparent Trojan signal ($2.2\times10^{-04}$), it would be discarded.  In addition, if any events were found to have durations inconsistent with the apparent Trojan signal they would also we discarded. The apparent Trojan eclipse has a duration of approximately $0.459$ days. Events would be deemed detections if they had durations of $0.459 \pm 0.05$ days.  This process was repeated for $10,000$ random periods. Figure \ref{fig:dimming_events} displays six examples of the detected dimming events as well as the apparent Trojan signal. Each window displays only part of the phase-folded residual as these detections often have durations much smaller than the phase as shown in Figure \ref{hist_detections}, dimming events similar to that associated with the hypothetical Trojan were more readily discovered at longer periods. While only four were detected around the orbital period of the Jovian, they were nevertheless detected making the systematic noise a serious candidate to explain the extra dimming.

\section{Discussion} \label{sec:disc}

Presented here is a re-analysis of the \emph{Kepler}-91 system. The results from the one-planet model for both photometry and radial velocity measurements re-confirm the planetary nature of the companion, \emph{Kepler}-91b, as a hot-Jupiter with characteristics similar to those estimated by \citet{Lillo-Box+etal:2014}.  Based on the model log-evidence calculations, \emph{Kepler}-91b is considerably more likely to be in an eccentric orbit, rather than circular, which is consistent with previous results. The estimate of the day-side temperature is consistent with the equilibrium temperature of the planet, although a possible confounding effect is that of the mid-eclipse brightening event that occurs during the secondary eclipse of \emph{Kepler}-91b. The predicted albedo is also consistent for short-period hot-Jupiters \citep{Angerhausen&Delarme:2014}.

The Trojan planet hypothesis was tested by applying a two-planet model (Planet+Trojan, eccentric) to the \emph{Kepler} photometric data that aims to describe \emph{Kepler}-91b and a hypothetical Trojan planet. This two-planet model was signicantly favored over the one-planet model by a Bayes' factor of   $\sim \exp(9) \approx 8000$, but was neither ruled out (nor verified) by the RV data obtained by \citet{Lillo-Box+etal:2014}.  
When performing Bayesian model testing, there are two important factors: the ratio of the prior probabilities of the models and the ratio of the evidences of the data.  When the prior probabilities of the models are assumed to be equal one can focus on the ratio of the evidences, or equivalently, the difference between their logarithms (the log odds ratio) \citep{Knuth+etal:2015}.  In the analysis presented here, we focused on the differences between the logarithms of the evidence, which implicitly assumes that the prior probabilities of the models are equal.  However, this assumption is questionable since our experience to date does not lead us to truly believe that it is as likely for \emph{Kepler}-91 to have a Jovian planet as it is for \emph{Kepler}-91 to have a Jovian planet with a Trojan companion.  The log evidences must be compared with this in mind.  Statistical evidence for the existence of Trojan planets has been uncovered \citep{Hippke&Angerhausen:2015}. However, given that at present there are about $5000$ exoplanet candidates and a Trojan companion to a planet has yet to be found, one could crudely estimate the ratio of probabilities to be something one the order of $1/5000$ against the Jovian+Trojan model, which corresponds to a log probability of about $6.9$.  Thus one would not be confident in such assertions unless the log evidence in support of a Trojan was proportionately higher. The fact that the log-evidence differences supporting a Trojan is on the order of $16$ (see Section \ref{res_phot}) yields an overall support with a probability on the order of $exp(16-6.9) \approx 9000$. 

Based on the model log-evidences for the single-planet case, we re-confirm the planetary nature of \emph{Kepler}-91b with an estimated minimum mass of $M_p \sin i = 0.86 \pm 0.09M_J$, which corresponds to a true mass of $M_p = 0.92 \pm 0.10M_J$ given the orbital inclination of $i = 69.8^o \pm 0.18^o$ determined using the photometric data.  This estimate of the mass of \emph{Kepler}-91b is consistent with the analysis of photometric data performed in Section \ref{res_phot}, along with the results from \citet{Lillo-Box+etal:2014}.  The eccentricity of the orbit was found to be $e = 0.041 \pm 0.028$, which also agrees with the photometric analysis within uncertainty.  Applying the two-planet (Planet + Trojan) model did not confirm or reject the Trojan hypothesis based on model log-evidences, however the two-planet model did have the lowest $\chi^2$ value, and highest log-likelihood making it a better fit to the data than the other models despite the Bayesian evidence being slightly lower than the one-planet eccentric model. 

\begin{figure}[h!]
\includegraphics[width=8cm]{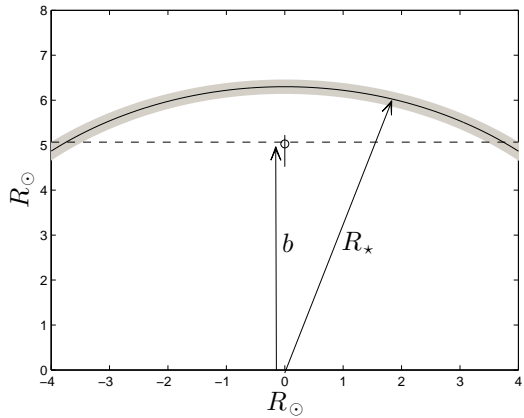}
\caption{Sky-projected separation between the centers of the planet and star.  The black arc represents the stellar limb ($R_\star = 6.30 \pm 0.16 R_\odot$) and the shaded regions represent the $\pm 1\sigma$ values for the stellar radius.  The planet is depicted as the small black circle and the vertical line through the planet represents the $\pm 1\sigma$ values for the impact parameter ($b = 5.06^{+0.19}_{-0.50}R_\odot$), and the dotted line is the sky-projected orbital path of the planet. }
\label{impact_parameter}
\end{figure}

The Jovian + Trojan model provides similar estimates of the \emph{Kepler}-91b parameters again supporting the idea that the planet is a hot Jupiter with mass $M_p = 0.99  \pm 0.07 M_J$, radius $R_p = 1.38 \pm 0.02R_J$, albedo $A_g = 0.39 \pm 0.17$ and day- and night-side temperatures of $2513.2 \pm 317.9$K and $2871.8 \pm 183.6$K, respectively.  The day- and night-side temperatures of \emph{Kepler}-91b in this model are equal to within uncertainty, and consistent with the expected equilibrium temperature of the planet suggesting that it may have high winds that equilibrate its day and night sides.  Results also suggest that a Trojan would be sub-Neptunian, with a radius $R_p = 2.91 \pm 0.56R_\oplus$, and with a mass of $M_p = 8.89 \pm 6.04 M_\oplus$.  The mass was not well estimated in this case because of the apparent lack of Doppler boosting and ellipsoidal variations associated with this object.  The day and night-side temperatures were determined to be a scorching $T_d = 5184.6 \pm 531.5$K and $T_n = 2372.6 \pm 1283.6$K, respectively.  While the uncertainty in these estimates are such that the temperatures are not conclusive, it is possible, according to the model, that the day-side of the planet may be greater than or approximately equal to the effective temperature of the host star, which has been determined to be $4550 \pm 75K$ \citep{Lillo-Box+etal:2013}).  The two-planet model fit illustrated in Figure \ref{fig:fit} shows the comparatively small depth of the Trojan transit to the corresponding secondary eclipse.  While this significant disparity in transit and secondary eclipse depths is apparent in the underlying data, and supportive of an extremely high day-side temperature, the particular values of the estimates could be affected by the mid-transit and mid-eclipse brightening events, which are neither modeled nor well-understood.  In addition, there are other known effects, which are not presently accounted for by EXONEST that could result in increased temperature estimates.  These include the extra illumination of the side of the planet opposite the star due to the planet's proximity (see for example, Figure 10 in \citet{Lillo-Box+etal:2013}) in addition to the high inclination of the orbit, which would allow part of the day-side of the planet to be visible from Earth during transit.  Also, a highly inclined and slightly eccentric orbit has a high probability to produce a grazing transit (see for example WASP-67; \citet{Hellier+etal:2012}).  However, using posterior estimates of the planetary radius and inclination (Table \ref{tbl:phot_parameters}) along with the published values of the stellar radius and semi-major axis (Table \ref{tbl:acceptedvalues}), one can calculate the estimated impact parameter, $b = 5.06^{+0.19}_{-0.50}R_\odot$.  Figure \ref{impact_parameter} displays the sky-projected distance between the center of the host star and the center of the planet, which shows that \emph{Kepler}-91b is most likely not in a grazing orbit.  

In addition, if a Trojan companion were present, both the Jovian and Trojan would undergo librations as discussed in Sections \ref{trojan hypothesis} and \ref{Trojan Stability Section}. Our simulations, based on the current model parameter estimates, show that the orbital phase of the Jovian planet would not significantly vary, whereas the Trojan's phase would vary.  One would expect that this would result in a smearing of the Trojan transit in the phase-folded light curve.  To properly model the system, N-body fitting routines should be employed rather than the Keplerian fitting used here.  While the libration periods obtained from our current model parameter estimates do not seem to be able to explain the observed even/odd period effects, a carefully modeling of the planet-planet interactions could account for additional unusual features observed in the light curve, such as the mideclipse brightening events.

Given the available data and the models employed, it is not yet possible to come to a conclusion as to the presence of a Trojan partner to \emph{Kepler}-91b.  In favor of the Trojan hypothesis is the fact that the Bayesian evidence of the Jovian+Trojan model is $exp(16)$ times greater than the Jovian model.  This hypothesis is still highly probable if one considers a reasonable prior probability reflecting the fact that a Trojan planet has never been observed in the set of $5000$ or so exoplanet candidates\footnote{Trojans have however been observed in our solar system at the Lagrange points of Venus, Earth, Mars,  Jupiter, Uranus, and Neptune.  In addition, Saturn's moon's Tethys and Dione each have two Trojan moons.}.  It is also remarkable that the model selected the appropriate relative phase for the Trojan companion.  While the probability of this occurring is not overwhelming, it is on the order of $1/36$.  However, this correct positioning of a Trojan occurred at the expense of having the secondary transit be deeper than the primary, which leads to the model ascribing an unphysically high day-side temperature to the Trojan, which clearly makes the Trojan hypothesis suspect.  In addition, unusual features, such as the odd/even phase differences in the light curve and the mid-eclipse brightening, which occurs not only during the Jovian eclipses, but also during hypothetical Trojan eclipses may be mimicking a Trojan-like signal.  At this stage, given the available data and the models employed, it is impossible to say anything definitive concerning the presence of a Trojan companion.

\section{Summary and Outlook}
We were able to confirm the planetary nature of \emph{Kepler}-91b using the EXONEST code for photometric analysis of its optical lightcurve obtained with the \emph{Kepler} Space telescope and radial velocity analysis on data taken from the literature.

We find that introducing an additional object to the system can explain the extra dimming in the lightcurve: an EXONEST photometric model including an object in the same orbit as \emph{Kepler}-91b, but shifted by $\sim 60$ degrees, produces a significantly higher Bayesian evidence than a model without this sub-Neptunian Trojan companion (see section \ref{res_phot}).  This model to describe the light curve is not ruled out by radial velocity, and the orbit seems to be stable over the course of 50,000 days, or 8000 cycles. On the other hand, this model also predicted an unphysical day-side temperature and a secondary eclipse depth greater than that of the transit, which would imply that either there are unknown heating mechanisms for such an object, or that it is a false-positive, which is far more likely.

 We are able to exclude previously suggested alternative explanations such as the presence of a moon, a resonant outer planet or instrumental effect for the observed dimmings caused by the  hypothetical  Trojan (see section \ref{dist_plan}, \ref{exo_moon} and \ref{instru}).  However, we detect a signficant odd/even effect in the phase-folded light curve, and similar dimming events in the light curve folded on randomized periods, both of which could mimic the hypothetical Trojan signal.


The \emph{Kepler}-91 system is currently of great interest since it is a short-period hot-Jupiter orbiting a star in its red giant stage.  If this Trojan planet is confirmed it would not only be the first detection of an exo-Trojan planet, but also would provide the unique opportunity to study two different worlds that are in identical stellar environments thus promising to provide insights into star-planet interactions. 

Despite this comprehensive investigation of the potential Trojan companion, given the data at hand we cannot conclude that it is the cause of the extra dimming events observed in the \emph{Kepler} light curve.

\section*{Acknowledgments}

The authors acknowledge the efforts of the \emph{Kepler} Mission team in obtaining the light curve data used in this publication. These data were generated by the \emph{Kepler} Mission science pipeline through the efforts of the \emph{Kepler} Science Operations Center and Science Office. The \emph{Kepler} Mission is lead by the project office at NASA Ames Research Center. Ball Aerospace built the \emph{Kepler} photometer and spacecraft which is operated by the mission operations center at LASP. These data products are archived at the Mikulski Archive for Space Telescopes.

The authors would thank the University at Albany Physics Department for its ongoing support in our exoplanet research.

Daniel Angerhausen's Research was supported by an appointment to the NASA Postdoctoral Program at the Goddard Space Flight Center, administered by Oak Ridge Associated Universities through a contract with NASA. 

The authors thank Em DeLarme, Jon Morse, William B. Rossow, and Barbara J. Thompson for discussions about this exciting extrasolar system.

%
\bibliographystyle{plainnat}
\bibliography{bibliography}

\end{document}